\documentclass[twocolumn]{aastex631}

\providecommand{\bjdtdb}{\ensuremath{\rm {BJD_{TDB}}}}

\providecommand{\msun}{\ensuremath{\,M_\Sun}}
\providecommand{\rsun}{\ensuremath{\,R_\Sun}}
\providecommand{\lsun}{\ensuremath{\,L_\Sun}}
\providecommand{\mj}{\ensuremath{\,M_{\rm J}}}
\providecommand{\rj}{\ensuremath{\,R_{\rm J}}}

\def\gtaprx{ \mathrel{ \vcenter{
      \offinterlineskip \hbox{$>$}
      \kern 0.3ex \hbox{$\sim$}    } } }
\def\ltaprx{ \mathrel{ \vcenter{
      \offinterlineskip \hbox{$<$}
      \kern 0.3ex \hbox{$\sim$}    } } }

\providecommand{\fave}{\langle F \rangle}
\providecommand{\fluxcgs}{10$^9$ erg s$^{-1}$ cm$^{-2}$}

\shorttitle{Revised Properties for the HD~17156 System}
\shortauthors{Stephen R. Kane et al.}

\begin{document}

\title{Revised Properties and Dynamical History for the HD~17156 System}

\author[0000-0002-7084-0529]{Stephen R. Kane}
\affiliation{Department of Earth and Planetary Sciences, University of
  California, Riverside, CA 92521, USA}
\email{skane@ucr.edu}

\author[0000-0002-0139-4756]{Michelle L. Hill}
\affiliation{Department of Earth and Planetary Sciences, University of
  California, Riverside, CA 92521, USA}

\author[0000-0002-4297-5506]{Paul A. Dalba}
\altaffiliation{Heising-Simons 51 Pegasi b Postdoctoral Fellow}
\affiliation{Department of Astronomy and Astrophysics, University of
  California, Santa Cruz, CA 95064, USA}
\affiliation{SETI Institute, Carl Sagan Center, 339 Bernardo Ave,
  Suite 200, Mountain View, CA 94043, USA}

\author[0000-0002-3551-279X]{Tara Fetherolf}
\altaffiliation{UC Chancellor's Fellow}
\affiliation{Department of Earth and Planetary Sciences, University of
  California, Riverside, CA 92521, USA}

\author[0000-0003-4155-8513]{Gregory W. Henry}
\affiliation{Center of Excellence in Information Systems, Tennessee
  State University, Nashville, TN 37209, USA}

\author[0000-0001-9309-0102]{Sergio B. Fajardo-Acosta}
\affiliation{Caltech/IPAC, Mail Code 100-22, Pasadena, CA 91125, USA}

\author[0000-0003-2519-6161]{Crystal~L.~Gnilka}
\affiliation{NASA Ames Research Center, Moffett Field, CA 94035, USA}

\author[0000-0001-8638-0320]{Andrew W. Howard}
\affiliation{Department of Astronomy, California Institute of
  Technology, Pasadena, CA 91125, USA}

\author[0000-0002-2532-2853]{Steve B. Howell}
\affiliation{NASA Ames Research Center, Moffett Field, CA 94035, USA}

\author[0000-0002-0531-1073]{Howard Isaacson}
\affiliation{Department of Astronomy, University of California,
  Berkeley, CA 94720, USA}
\affiliation{Centre for Astrophysics, University of Southern
  Queensland, Toowoomba, QLD 4350, Australia}


\begin{abstract}

From the thousands of known exoplanets, those that transit bright host
stars provide the greatest accessibility toward detailed system
characterization. The first known such planets were generally
discovered using the radial velocity technique, then later found to
transit. HD~17156b is particularly notable among these initial
discoveries because it diverged from the typical hot Jupiter
population, occupying a 21.2~day eccentric ($e = 0.68$) orbit,
offering preliminary insights into the evolution of planets in extreme
orbits. Here we present new data for this system, including ground and
space-based photometry, radial velocities, and speckle imaging, that
further constrain the system properties and stellar/planetary
multiplicity. These data include photometry from the Transiting
Exoplanet Survey Satellite ({\it TESS}) that cover five transits of
the known planet. We show that the system does not harbor any
additional giant planets interior to 10~AU. The lack of stellar
companions and the age of the system indicate that the eccentricity of
the known planet may have resulted from a previous planet-planet
scattering event. We provide the results from dynamical simulations
that suggest possible properties of an additional planet that
culminated in ejection from the system, leaving a legacy of the
observed high eccentricity for HD~17156b.

\end{abstract}



\section{Introduction}
\label{intro}

The large number of exoplanet discoveries have uncovered a diverse
range of planetary architectures, many of which differ significantly
from the planets and orbits found in the solar system
\citep{ford2014,winn2015,kane2021d}. One of the more extreme
divergences from the solar system architecture is that of highly
eccentric giant planets within the broader eccentricity distribution
\citep{shen2008c,kane2012d,vaneylen2015}. The existence of eccentric
giant planets may be the result of disk interactions during formation
\citep{clement2021e} or may reveal a potentially dynamically turbulent
past regarding planet-planet scattering events
\citep{chatterjee2008,ford2008c,kane2014b,carrera2019b}. Such planets
are extremely important with respect to the evolution of planetary
system dynamics \citep{juric2008b,ford2014,winn2015}, including the
possible locations of potentially habitable terrestrial planets in the
system
\citep{kane2012e,georgakarakos2018,hill2018,sanchez2018,kane2019e}. Highly
eccentric planets also provide opportunities to study atmospheric
circulation and radiative forcing in extreme flux environments
\citep{kane2011g,kataria2013,lewis2013a}. Eccentric planets that
transit their host star are therefore particularly valuable assets in
the exoplanet inventory since they reveal mass-radius relations and
atmospheric information via transmission spectroscopy
\citep{kane2009b,mayorga2021a}. Fortunately, eccentric planets also
have an enhanced transit probability
\citep{barnes2007d,burke2008a,kane2008b}, resulting in several key
discoveries of long-period transiting planets in eccentric orbits,
such as HD~80606b \citep{naef2001b,laughlin2009a,dewit2016a}, and the
more recent case of Kepler-1704b \citep{dalba2021c}.

Among the early detection of planetary transits, the most significant
were those planets discovered with the radial velocity (RV) method,
due to the relative brightness of their host stars
\citep{kane2007b,kane2009c}. The Transiting Exoplanet Survey Satellite
({\it TESS}) has carried out photometric monitoring of bright stars
throughout the sky since its launch in 2018 \citep{ricker2015},
including many known RV exoplanet host stars
\citep{dalba2019c,kane2021b}. These {\it TESS} observations have
enabled the detection of transits for several systems, including
HD~118203 \citep{pepper2020} and HD~136352 \citep{kane2020c}. Prior to
the launch of the {\it Kepler} mission, the limited group of RV
transiting planets included the very first detected transiting planet;
HD~209458b \citep{charbonneau2000,henry2000a}. A significant milestone
planet is HD~17156b, a Jovian-mass planet discovered via RVs by
\citet{fischer2007b} and then subsequently found to transit the host
star by \citet{barbieri2007}. The high interest in the planet stemmed
largely from the divergence from previous hot Jupiter discoveries,
both in terms of its relatively large orbital period ($P = 21.2$~days)
and eccentricity ($e = 0.68$). The interest in the system resulted in
numerous follow-up observations to refine the system parameters,
including the planet size/mass and orbit
\citep{gillon2008a,irwin2008d,barbieri2009,winn2009a,dawson2012b} and
space-based observations to characterize the host star and
star--planet interactions
\citep{gilliland2011a,nutzman2011a,southworth2011e,maggio2015}. The
HD~17156 system was also intensively studied with respect to the
potential misalignment between the planetary orbital axis and the
stellar rotational axis via detection of the Rossiter–McLaughlin (R-M)
effect. The spin-orbit was initially found to exhibit substantial
misalignment \citep{narita2008}, but further observations indicated
that the spin-orbit misalignment was small
\citep{cochran2008a,narita2009a}. With all of these follow-up
observations, the HD~17156 system remains a crucial milestone in our
knowledge of eccentric orbits, giant planet formation and evolution,
and orbital dynamics within planetary systems.

In this paper, we present new observations of HD~17156, including
ground-based photometry to measure the long-term stellar variability,
{\it TESS} photometry that reveals five transits of the known
exoplanet, new RV data that updates the orbit, and speckle imaging to
constrain the presence of stellar companions. These observations
improve the mass, radius, and orbit of the planet, and provide insight
into the dynamical history of the planet. Section~\ref{obs} describes
the data sources for the photometric, RV, and imaging components of
the observations. Section~\ref{results} presents the results from the
data analysis, including revised properties for both the star and
planet, and constraints on additional bodies within the system. We
discuss the implications of our data and analysis for the dynamical
history of the planet in Section~\ref{dynamical}, then provide
concluding remarks and suggestions for further work in
Section~\ref{conclusions}.


\section{Observations}
\label{obs}

Due to the long-standing interest in HD~17156 (HIP~13192,
TIC~302773669, TOI-1573), the star has been observed on numerous
occasions. Here, we describe observations carried out for this
project, combining space and ground-based photometry, additional RVs,
and speckle imaging.


\subsection{Photometry}
\label{photometry}


\subsubsection{T12 APT}
\label{apt}

We acquired 1059 out-of-transit photometric observations of HD~17156
over 10 observing seasons from 2006--07 to 2016--17. The data were
acquired with the T12 0.80~m automatic photoelectric telescope (APT)
at Fairborn Observatory in Arizona. The T12 APT has a dual channel
photometer equipped with two EMI 9124QB photomultiplier tubes to
measure differential magnitudes simultaneously in the Str\"omgren $b$
and $y$ passbands. To improve the photometric precision of the
individual nightly observations, we combine the differential $b$ and
$y$ magnitudes into a single $(b+y)/2$ ``passband.'' The precision of
a single observation with T12, as measured from pairs of constant
comparison stars, is typically around 0.0015--0.0020~mag on good
nights. The T12 APT is functionally identical to the T8 APT, described
in detail by \citet{henry1999}.

\begin{figure}
  \includegraphics[width=8.5cm]{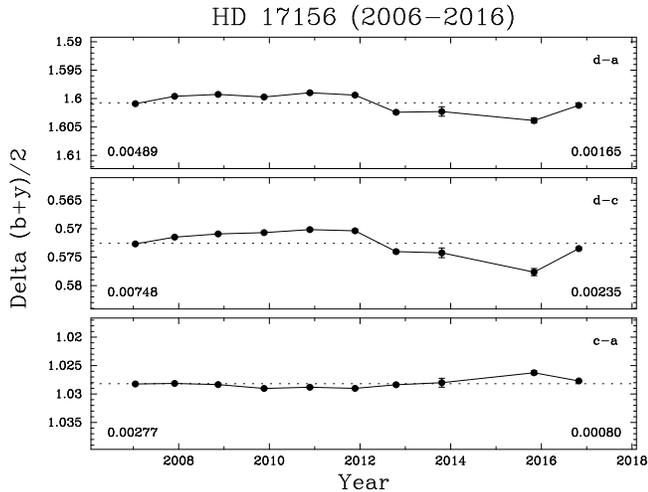}
  \caption{Comparison of the seasonal means of the d-a, d-c, and c-a
    differential magnitudes show that the observed variability is
    intrinsic to HD~17156.}
  \label{fig:apt1}
\end{figure}

The comparison stars were HD~15784 (star a), HD~19016 (star b), HD
16066 (star c), with HD~17156 designated as star d. Intercomparison of
the 6 combinations of differential magnitudes (d-a, d-b, d-c, c-a,
c-b, b-a) shows that star b is a low amplitude variable, while stars a
and c are constant to the limit of our precision. Therefore, we
created differential magnitudes in the sense HD~17156 minus the mean
brightness of HD~15784 and HD~16066 in the combined $(b+y)/2$
passband. Figure~\ref{fig:apt1} plots the seasonal mean differential
magnitudes of d-a, d-c, and c-a in the top, middle, and bottom panels,
respectively. The numbers in the lower right of each panel are the
standard deviations of the mean magnitudes from the mean of the mean
magnitudes, represented by the dotted line in each panel. The numbers
in the lower left give the total range in the seasonal
means. Comparison of the three panels shows that most of the
variability detected in these three stars is intrinsic to HD~17156.

Table~\ref{tab:apt} summarizes the observations of the d-ac
differential magnitudes of HD~17156. Most standard deviations of the
nightly observations from their individual seasonal means (column~4)
fall between 0.00134 and 0.00212 mag, so the night-to-night scatter in
the observations is similar to the typical measurement
uncertainty. Table~\ref{tab:apt} also lists the 10 seasonal means
(column~5), along with the last two digits of their standard
deviations. The seasonal means cover a range of 0.00618 mag.

\begin{deluxetable*}{ccccc}
  \tablewidth{0pc}
  \tablecaption{\label{tab:apt} Summary of T12 APT photometric
    observations for HD~17156.}
  \tablehead{
\colhead{Observing} & \colhead{} & \colhead{Date Range} & \colhead{Sigma} & 
\colhead{Seasonal Mean} \\
\colhead{Season} & \colhead{$N_{obs}$} & \colhead{(HJD $-$ 2400000)} & 
\colhead{(mag)} & \colhead{(mag)}
  }
  \startdata
2006--07   & 214 & 54001--54179 & 0.00212 & 1.08680(15) \\
2007--08   & 400 & 54370--54535 & 0.00183 & 1.08554(09) \\
2008--09   & 124 & 54728--54881 & 0.00146 & 1.08509(13) \\
2009--10   &  64 & 55092--55245 & 0.00141 & 1.08520(18) \\
2010--11   &  69 & 55463--55610 & 0.00160 & 1.08457(19) \\
2011--12   &  64 & 55823--55971 & 0.00154 & 1.08488(19) \\
2012--13   &  35 & 56186--56265 & 0.00148 & 1.08821(25) \\
2013--14   &  12 & 56558--56634 & 0.00259 & 1.08826(75) \\
2014--15   &   0 &   \nodata    & \nodata &   \nodata   \\
2015--16   &  33 & 57293--57377 & 0.00294 & 1.09075(51) \\
2016--17   &  44 & 57666--57734 & 0.00134 & 1.08734(20) \\
  \enddata
\end{deluxetable*}

Analysis of the nightly d-ac differential magnitudes finds no
significant periodicity between 1 and 100 days within any of the 10
individual observing seasons or in the data set as a whole. In
particular, we find no evidence for periodicity around the estimated
17.8~day rotation period discussed in Section~\ref{system}. Therefore,
rotational modulation in the brightness of HD~17156 due to star spots
is undetectable in our photometric observations. Furthermore, we find
no significant variability at or around the 21.2~day period of the
known planet, providing further evidence that the RV variations are
due to the planetary reflex motion of the star.

\begin{figure}
  \includegraphics[angle=270,width=8.5cm]{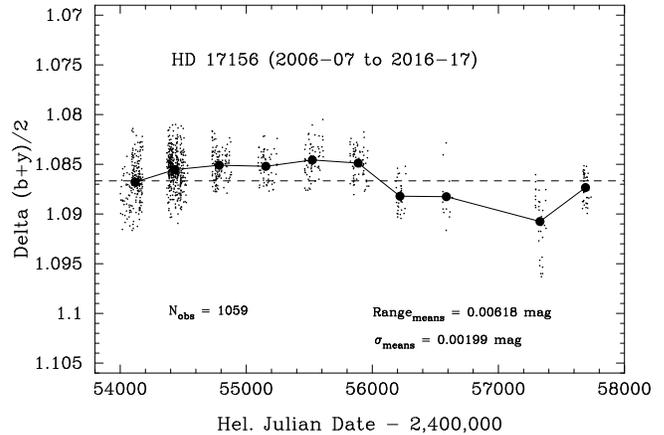}
  \caption{Nightly Str\"omgren $(b+y)/2$ band photometry of HD~17156
    from 10 observing seasons from 2006-07 to 2016-17 (small circles),
    acquired with the T12 0.80m APT at Fairborn Observatory. The star
    is constant from night-to-night within most observing seasons to
    the limit of our precision. The seasonal mean magnitudes are
    plotted as the large filled circles and cover a range of
    0.00618~mag with a standard deviation from the mean of the
    seasonal means of 0.00199~mag, indicating low-amplitude
    year-to-year variability in HD~17156. The seasonal means suggest a
    stellar cycle of around 10 years.}
  \label{fig:apt2}
\end{figure}

Figure~\ref{fig:apt2} plots the individual nightly differential
magnitudes from the 10 observing seasons as small filled circles. The
seasonal means are plotted as the large filled circles. The standard
deviations of the individual seasonal means (see Table~\ref{tab:apt})
are roughly the size of the plot symbols. The standard deviation of
the individual mean magnitudes from their grand mean is 0.00199~mag,
which is several times larger than the standard deviation of the
individual means. The mean magnitudes in Figure~\ref{fig:apt2} suggest
a stellar cycle in HD~17156 of $\sim$10~years.


\subsubsection{TESS}
\label{tess}

The {\it TESS} spacecraft observed HD~17156 during Sector 18
(2019-Nov-02 to 2019-Nov-27, in cycle 2), Sector 19 (2019-Nov-27 to
2019-Dec-24, in cycle 2), Sector 25 (2020-May-13 to 2020-Jun-08, in
cycle 2), and Sector 52 (2022-May-18 to 2022-Jun-13, in cycle
4). These {\it TESS} data can be found in MAST:
\dataset[10.17909/t9-nmc8-f686]{https://doi:10.17909/t9-nmc8-f686}. HD~17156
is relatively bright ($V \sim 8.2$) and so was observed with two
minute cadence, compared to the 30 minute sampling for most of the
sky. Since the orbital period of HD~17156b ($P = 21.2$~days) is less
than the typical dwell time for {\it TESS} on a given sector, a
transit event is all but guaranteed during a particular
sector. Indeed, a single transit was detected in each of the Sectors
18, 19, and 25, and two transits were observed during Sector 52. The
transits were easily detected by the Science Processing Operations
Center (SPOC) pipeline, so HD~17156 was assigned {\it TESS} Object of
Interest (TOI) number 1573.

\begin{figure*}
  \centering
  \includegraphics[width=16.0cm]{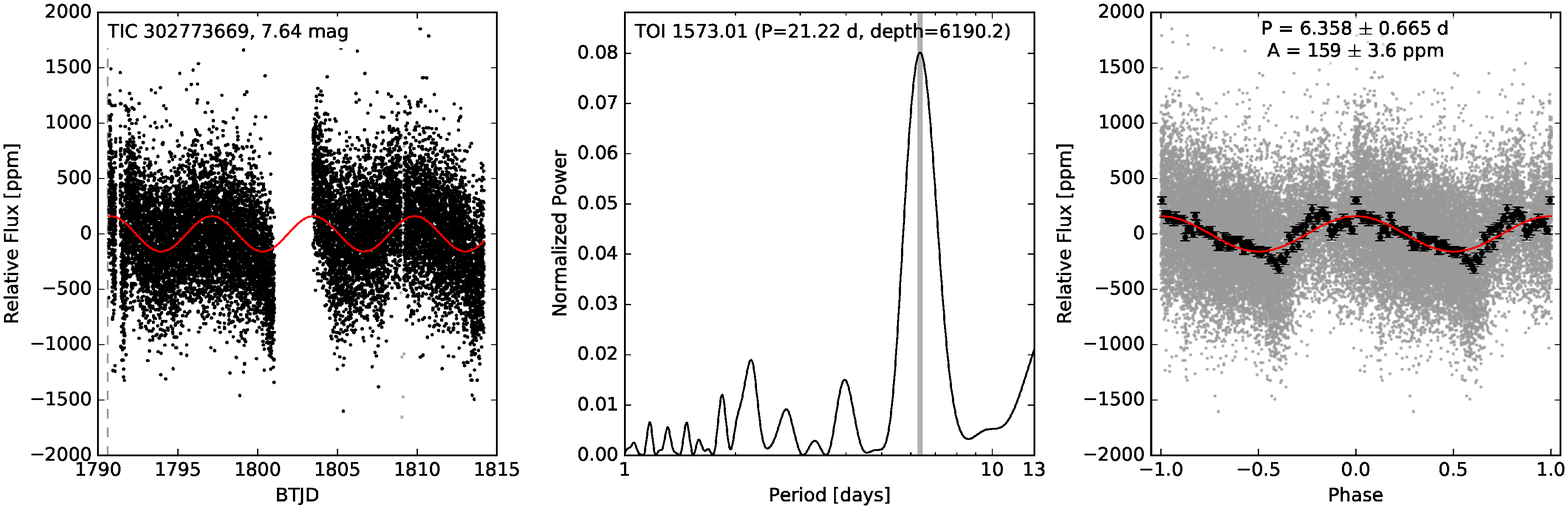} \\
  \includegraphics[width=16.0cm]{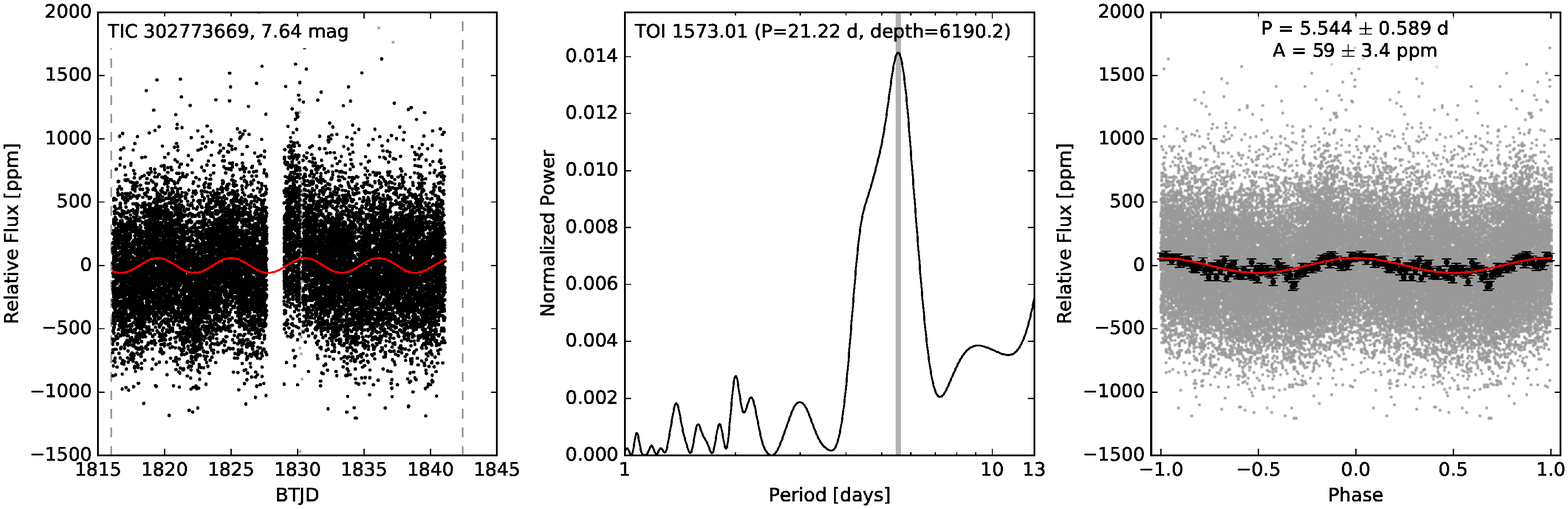} \\
  \includegraphics[width=16.0cm]{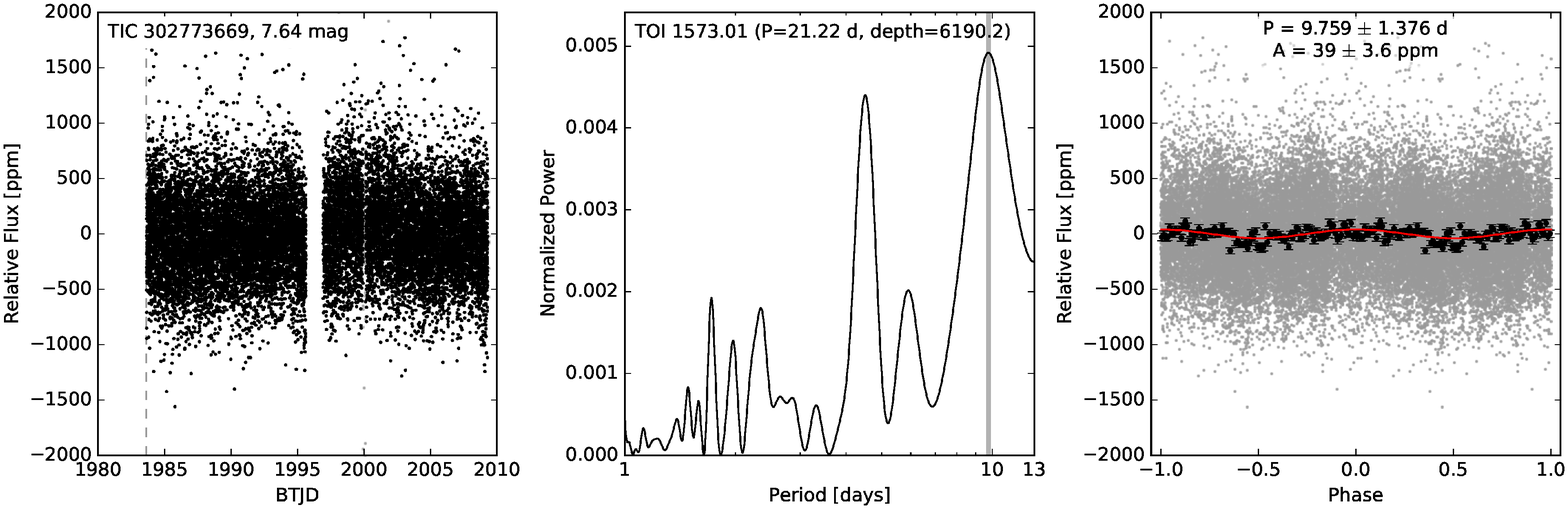} \\
  \includegraphics[width=16.0cm]{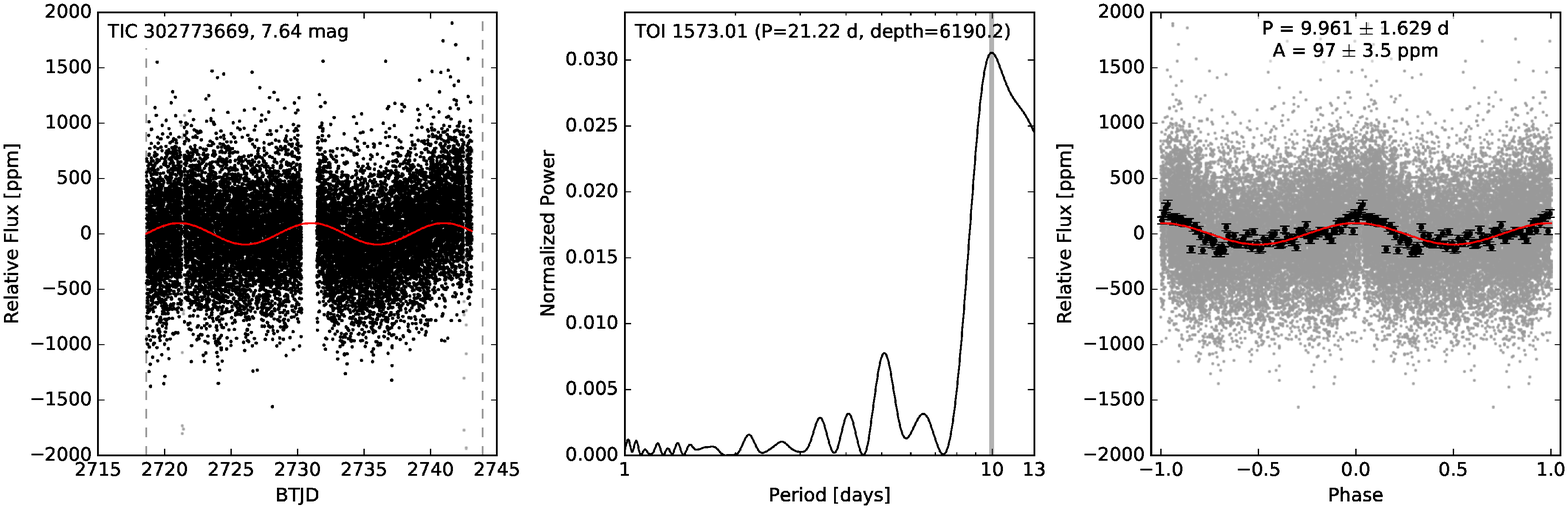}
  \caption{The {\it TESS} 2-min PDC light curves (left), LS
    periodograms (center), and phase-folded light curves (right) from
    the variability analysis for HD~17156. Dates in the left panel are
    shown in Barycentric {\it TESS} Julian Date (BTJD). The
    variability analysis is performed separately on each {\it TESS}
    sector, which is ordered in time from top to bottom: Sectors 18,
    19, 25, and 52. The mean out-of-transit photometric scatter over
    all sectors is 380~ppm. The red curve shows a sinusoidal fit to
    the most significant periodicity. The gray points in the left
    panel indicate data that were removed from the variability
    analysis, which includes data flagged as poor quality, 5$\sigma$
    outliers, and transits of HD~17156b. In the right panel, the gray
    points indicate all data included in the variability analysis and
    the black points represent the binned data. The periodic signals
    detected by the LS periodograms are inconsistent between sectors
    and are low in normalized power ($< 0.1$), such that we consider
    HD 17156 to be a quiet, non-variable star on timescales $< 13$
    days.}
  \label{fig:tess}
\end{figure*}

Prior to our transit analysis, we investigated the variability of
HD~17156 using the methodology described in \citet{fetherolf2022}. In
brief, a Lomb-Scargle \citep[LS;][]{lomb1976,scargle1982} periodogram
search from 1--13 days was performed on each sector of the 2-min
cadence Pre-search Data Conditioning (PDC) photometry acquired by {\it
  TESS}. A careful vetting process for determining significant
periodic stellar variability and excluding systematic aliases was
performed by \citet{fetherolf2022}, and the resulting variability
catalog does not include HD~17156. Figure~\ref{fig:tess} shows the
light curve (left), LS periodogram (center), and phase-folded light
curve (right) for each {\it TESS} sector of photometry available for
HD~17156 at the time of writing. The detected periodic signatures are
inconsistent between {\it TESS} sectors and none of the LS
periodograms reach a high-significance threshold of 0.1 normalized
power. The most significant periodicity is Sector 18 (0.08 normalized
power), but the periodogram is clearly biased towards upwards and
downwards trends that occur near the spacecraft's data uplink
times. Therefore, we consider HD~17156 to be a relatively quiet,
non-variable star on timescales of $<$13 days, in agreement with the
APT results described in Section~\ref{apt}.

In preparation for the transit analysis, the {\it TESS} data were
detrended for stellar variability and instrumental effects through the
use of the
keplersplinev2\footnote{\url{https://github.com/avanderburg/keplersplinev2}}
tool \citep{vanderburg2014}. The analysis of the planetary signatures
in these data are described in Section~\ref{results}.


\subsection{Radial Velocities}
\label{rv}

HD~17156 has been observed using the HIRES echelle spectrometer
\citep{vogt1994} on the 10m Keck I telescope since early 2006. The
continued observations occurred within the framework of the California
Legacy Survey, described in more detail by
\citet{rosenthal2021,fulton2021,rosenthal2022}. The Keck RV
measurements were created from observations with an iodine cell,
producing a dense set of molecular absorption lines imprinted on the
stellar spectra that enable robust wavelength calibration from which
precision Doppler measurements and instrumental profile constraints
are produced \citep{marcy1992a,valenti1995b}. The Doppler shift for
each star-times-iodine spectrum were extracted using the modeling
techniques described by \citep{butler1996b,howard2009a}. A subset of
the RV data, including times of observation, relative RVs, and
associated errors for the Keck data, are shown in
Table~\ref{tab:rv}. In total, 71 RV measurements are included in our
full dataset that span a period of $\sim$17~years, with a median
uncertainty of 1.415~m/s. Note that there are other RV data sources
that focus on the specific orbital location of inferior conjunction
for the purpose of detecting the R-M effect
\citep{cochran2008a,narita2008,narita2009a}. However, those data are
not included in this analysis as we found that they do not contribute
significantly to the overall Keplerian orbital fit, which is dominated
by the long-term nature of the Keck/HIRES data.

\begin{deluxetable}{ccc}
  \tablewidth{0pc}
  \tablecaption{\label{tab:rv} HD~17156 radial velocities.}
  \tablehead{
    \colhead{Date} &
    \colhead{RV} &
    \colhead{$\sigma$} \\
    \colhead{(BJD -- 2450000)} &
    \colhead{(m/s)} &
    \colhead{(m/s)}
  }
  \startdata
3746.7593 & -15.279 & 1.639 \\
3748.8014 & 31.516 & 1.776 \\
3749.7980 & 42.631 & 1.780 \\
3750.8048 & 63.699 & 1.569 \\
3775.7800 & 130.727 & 1.784 \\
3776.8097 & 150.371 & 1.628 \\
3779.8306 & 133.042 & 1.563 \\
3959.1318 & -5.680 & 1.415 \\
3962.0700 & 47.436 & 1.210 \\
3963.1059 & 62.608 & 1.586 \\
3964.1310 & 92.630 & 1.532 \\
3982.0333 & 29.756 & 1.074 \\
3983.0868 & 41.970 & 1.544 \\
3983.9959 & 61.650 & 1.243 \\
3985.0096 & 85.931 & 1.512 \\
4023.9553 & 5.941 & 1.834 \\
4047.9618 & 62.277 & 1.799 \\
4083.9073 & -69.172 & 1.226 \\
4084.8328 & -38.746 & 1.396 \\
4085.8695 & -20.385 & 1.650 \\
4129.9276 & 9.324 & 1.401 \\
4130.7326 & 30.847 & 1.451 \\
4131.8572 & 43.177 & 1.900 \\
4138.7692 & 162.379 & 1.311 \\
4319.1285 & -18.306 & 1.146 \\
4336.0806 & -156.118 & 1.169 \\
4337.1220 & -117.681 & 1.357 \\
4339.1313 & -47.422 & 1.196 \\
4427.8273 & 31.948 & 1.419 \\
4428.8656 & 51.232 & 1.529 \\
4545.7235 & -361.921 & 1.387 \\
4545.7276 & -364.986 & 1.285 \\
4546.8283 & -254.029 & 1.481 \\
4546.8339 & -254.519 & 1.435 \\
4673.1254 & -356.450 & 1.251 \\
4702.1282 & 8.626 & 1.095 \\
4703.0358 & 27.300 & 1.101 \\
4704.1246 & 41.815 & 1.172 \\
4705.0540 & 60.832 & 1.227 \\
  \enddata
\tablenotetext{}{The full data set is available online.}
\end{deluxetable}


\subsection{Speckle Imaging}
\label{imaging}

\begin{figure}
  \includegraphics[width=8.5cm]{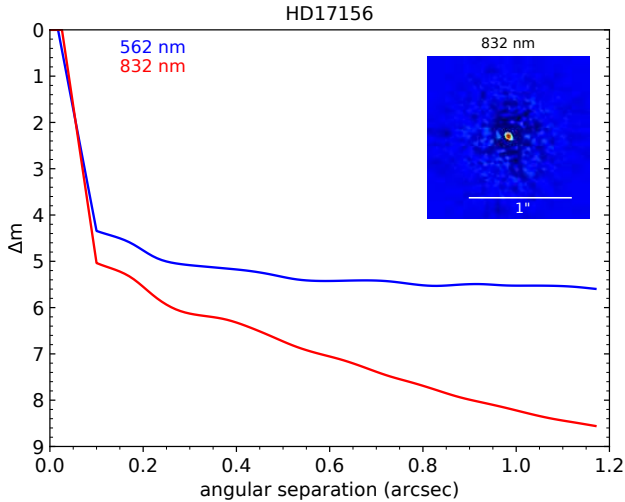}
  \caption{The 562nm (blue) and 832nm (red) contrast curve results and
    reconstructed speckle image for HD~17156, from observations
    carried out using the 'Alopeke instrument at the Gemini North
    Observatory.}
  \label{fig:imaging}
\end{figure}

If a star hosting a planet candidate has a close bound companion (or
companions), the companion can create a false-positive exoplanet
detection if it is an eclipsing binary (EB). Additionally, flux from
the additional source(s) can lead to an underestimated planetary
radius if not accounted for in the transit model
\citep{ciardi2015a,matson2018}. In order to ascertain the possibility
of close, low-mass stellar companions to HD~17156, we conducted
observations using the 'Alopeke instrument at the Gemini North
Observatory \citep{scott2021b}. 'Alopeke provides simultaneous speckle
imaging in two bands (562nm and 832nm) with output data products
including a reconstructed image and robust contrast limits on
companion detections \citep{howell2011}. The observations were carried
out on 2022 September 14. No apparent sign of a stellar companion was
detected from the imaging data down to the sensitivity limit of the
instrument. Figure~\ref{fig:imaging} shows our 562nm and 832nm
contrast curve results and our reconstructed speckle
image. Specifically, we find that HD~17156 is a single star with no
close companion brighter than 5 to 8.5 magnitudes within the 5-sigma
contrast and angular limits achieved (0.02\arcsec--1.2\arcsec). The
angular limits, at the distance of HD~17156 ($d = 78$~pc; see
Table~\ref{tab:star}), correspond to spatial limits of 1.6~AU to
94~AU. These results are consistent with those from
\citet{adams2013c}, who did not detect any stellar companions within
the angular range of 0.5\arcsec--4.0\arcsec (39--312~AU). The
implications of these results are discussed further in
Section~\ref{dynamical}.

\section{Results}
\label{results}

The data described in Section~\ref{obs} provide a solid foundation
from which to construct a thorough analysis and characterization of
the star, planet, and other possible companions within the system.


\subsection{Extraction of System Parameters}
\label{system}

Here we provide new and updated properties for the HD~17156
system. The extraction of stellar and planetary properties was largely
performed using the {\sc EXOFASTv2}
tool\footnote{\url{https://github.com/jdeast/EXOFASTv2}}, described in
detail by \citet{eastman2013,eastman2020}. We followed a similar
application of {\sc EXOFASTv2} to that used by \citet{kane2020c},
whilst applying noise floors to the stellar effective temperature and
bolometric flux \citep{tayar2022a}. We derived stellar properties for
our sample by applying the {\sc SpecMatch} \citep{petigura2015} and
{\sc Isoclassify} \citep{huber2017d} software packages to the template
Keck-HIRES spectra of our stars. {\sc Specmatch} takes an optical
stellar spectrum as input, and by interpolating over a grid of
template spectra with known associated stellar properties, returns
three spectral properties and uncertainties. We updated the normal
prior on parallax to those provided by the third data release of the
{\it Gaia} mission \citep{brown2021}, including the corrections
provided by \citet{lindegren2021b}. Convergence for the global fit to
the transit and RV data was assessed using the default {\sc EXOFASTv2}
statistics of $T_z$ \citep{ford2006d}, the number of independent draws
of the underlying posterior probability distribution (convergence for
$T_z>1000$ for each parameter), and GR, the Gelman-Rubin statistic
\citep{gelman1992}, where convergence is achieved for GR$<1.01$ for
each parameter.

\begin{deluxetable*}{lcc}
\tablecaption{\label{tab:star} HD~17156 derived stellar parameters.}
\tablehead{\colhead{~~~Parameter} & \colhead{Units} & \colhead{Values}}
\startdata
~~~~$M_*$\dotfill &Mass (\msun)\dotfill &$1.285^{+0.064}_{-0.062}$\\
~~~~$R_*$\dotfill &Radius (\rsun)\dotfill &$1.517^{+0.038}_{-0.036}$\\
~~~~$L_*$\dotfill &Luminosity (\lsun)\dotfill &$2.76^{+0.19}_{-0.13}$\\
~~~~$F_{Bol}$\dotfill &Bolometric Flux (cgs)\dotfill &$0.00000001477^{+0.00000000100}_{-0.00000000071}$\\
~~~~$\rho_*$\dotfill &Density (cgs)\dotfill &$0.517^{+0.039}_{-0.035}$\\
~~~~$\log{g}$\dotfill &Surface gravity (cgs)\dotfill &$4.184\pm0.024$\\
~~~~$T_{\rm eff}$\dotfill &Effective Temperature (K)\dotfill &$6046^{+76}_{-72}$\\
~~~~$[{\rm Fe/H}]$\dotfill &Metallicity (dex)\dotfill &$0.208\pm0.058$\\
~~~~$Age$\dotfill &Age (Gyr)\dotfill &$3.3^{+1.2}_{-1.0}$\\
~~~~$A_V$\dotfill &V-band extinction (mag)\dotfill &$0.090^{+0.082}_{-0.061}$\\
~~~~$\varpi$\dotfill &Parallax (mas)\dotfill &$12.941\pm0.039$\\
~~~~$d$\dotfill &Distance (pc)\dotfill &$77.27\pm0.23$\\
\smallskip\\\multicolumn{2}{l}{Wavelength Parameters:}&TESS\smallskip\\
~~~~$u_{1}$\dotfill &linear limb-darkening coeff \dotfill &$0.262\pm0.019$\\
~~~~$u_{2}$\dotfill &quadratic limb-darkening coeff \dotfill &$0.289\pm0.022$\\
\enddata
\end{deluxetable*}

The derived stellar parameters from the global {\sc EXOFASTv2} fit are
shown in Table~\ref{tab:star}. In summary, HD~17156 is a G0 sub-giant
star, slightly more massive than the Sun, and with an age of
$\sim$3.3~Gyr. In addition to these parameters, {\sc SpecMatch}
analysis provided a projected stellar rotational velocity of $v \sin i
= 4.32\pm1.0$~km/s. As described in Section~\ref{intro}, analysis of
the R-M effect via RV data of the system during planetary transit
revealed that the system exhibits a relatively small spin-orbit
misalignment. Thus the projected stellar rotational velocity is a good
approximation for the true rotational velocity, which predicts a
rotation period of $\sim$17.8~days, consistent with the above cited
stellar age. As described in Section~\ref{obs}, we do not detect
evidence of stellar variability on short timescales, including any
periodic signals near 17~days. However, the APT photometry
(Section~\ref{apt}) indicates the presence of a $\sim$10~year
photometric signature, possibly due to the the magnetic activity cycle
of the host star \citep{strassmeier2005a,dragomir2012a}.

The planetary parameters derived from the {\sc EXOFASTv2} analysis are
provided in Table~\ref{tab:planet}. There are numerous items of note
regarding the data in this table. Timing information are shown with
the subscript ``TDB'', which is the Barycentric Dynamical Time, which
includes relativistic corrections that move the origin to the
barycenter. The revised orbital period has exceptionally small
uncertainties similar to that determined by \citep{ivshina2022},
though our fit includes more transits and the combination with the RV
data. The equilibrium temperature ($\sim$888~K) is calculated at the
semi-major axis of the orbit and assumes no albedo and perfect heat
redistribution \citep{kane2011g}. Using these same assumptions, the
equilibrium temperature approaches 1600~K during periastron
passage. The eclipse impact parameter, $b_S$, is greater than unity,
since the eccentricity and periastron argument of the orbit ensure
that the planet does not pass directly behind the host star during
superior conjunction. $V_c/V_e$ is the velocity ratio of the planet
between circular and eccentric orbit scenarios during inferior
conjunction, indicating the significant reduction in transit duration
caused by the orbital orientation relative to the line of sight.

\begin{deluxetable*}{lcc}
\tablecaption{\label{tab:planet} HD~17156 planetary parameters.}
\tablehead{\colhead{~~~Parameter} & \colhead{Units} & \colhead{Values}}
\startdata
~~~~$P$\dotfill &Period (days)\dotfill &$21.2164294\pm0.0000061$\\
~~~~$R_p$\dotfill &Radius (\rj)\dotfill &$1.094^{+0.031}_{-0.030}$\\
~~~~$M_p$\dotfill &Mass (\mj)\dotfill &$3.26\pm0.11$\\
~~~~$T_C$\dotfill &Time of conjunction (\bjdtdb)\dotfill &$2458809.07037\pm0.00021$\\
~~~~$a$\dotfill &Semi-major axis (AU)\dotfill &$0.1632\pm0.0027$\\
~~~~$i$\dotfill &Inclination (Degrees)\dotfill &$86.51^{+0.37}_{-0.34}$\\
~~~~$e$\dotfill &Eccentricity \dotfill &$0.6772^{+0.0045}_{-0.0044}$\\
~~~~$\omega_*$\dotfill &Argument of Periastron (Degrees)\dotfill &$122.06\pm0.37$\\
~~~~$T_{\rm eq}$\dotfill &Equilibrium temperature (K)\dotfill &$888^{+12}_{-11}$\\
~~~~$\tau_{\rm circ}$\dotfill &Tidal circularization timescale (Gyr)\dotfill &$20.7^{+3.8}_{-3.2}$\\
~~~~$K$\dotfill &RV semi-amplitude (m/s)\dotfill &$274.5^{+2.5}_{-2.3}$\\
~~~~$R_p/R_*$\dotfill &Radius of planet in stellar radii \dotfill &$0.07412^{+0.00039}_{-0.00040}$\\
~~~~$a/R_*$\dotfill &Semi-major axis in stellar radii \dotfill &$23.11^{+0.56}_{-0.53}$\\
~~~~$\delta$\dotfill &$\left(R_P/R_*\right)^2$ \dotfill &$0.005493^{+0.000058}_{-0.000059}$\\
~~~~$\delta_{\rm TESS}$\dotfill &Transit depth in TESS (fraction)\dotfill &$0.006085\pm0.000055$\\
~~~~$\tau$\dotfill &Ingress/egress transit duration (days)\dotfill &$0.01162^{+0.00062}_{-0.00060}$\\
~~~~$T_{14}$\dotfill &Total transit duration (days)\dotfill &$0.13127^{+0.00066}_{-0.00064}$\\
~~~~$T_{\rm FWHM}$\dotfill &FWHM transit duration (days)\dotfill &$0.11965^{+0.00036}_{-0.00035}$\\
~~~~$b$\dotfill &Transit Impact parameter \dotfill &$0.484^{+0.035}_{-0.041}$\\
~~~~$b_S$\dotfill &Eclipse impact parameter \dotfill &$1.79^{+0.13}_{-0.15}$\\
~~~~$\rho_p$\dotfill &Density (cgs)\dotfill &$3.08^{+0.26}_{-0.23}$\\
~~~~$\log{g_p}$\dotfill &Surface gravity \dotfill &$3.829\pm0.024$\\
~~~~$\Theta$\dotfill &Safronov Number \dotfill &$0.757\pm0.021$\\
~~~~$\fave$\dotfill &Incident Flux (\fluxcgs)\dotfill &$0.0937^{+0.0053}_{-0.0044}$\\
~~~~$T_P$\dotfill &Time of Periastron (\bjdtdb)\dotfill &$2458788.1332^{+0.0056}_{-0.0055}$\\
~~~~$T_S$\dotfill &Time of eclipse (\bjdtdb)\dotfill &$2458813.946^{+0.080}_{-0.084}$\\
~~~~$V_c/V_e$\dotfill & \dotfill &$0.4675^{+0.0034}_{-0.0035}$\\
~~~~$e\cos{\omega_*}$\dotfill & \dotfill &$-0.3595^{+0.0051}_{-0.0053}$\\
~~~~$e\sin{\omega_*}$\dotfill & \dotfill &$0.5739^{+0.0035}_{-0.0034}$\\
~~~~$M_p\sin i$\dotfill &Minimum mass (\mj)\dotfill &$3.26\pm0.11$\\
~~~~$M_p/M_*$\dotfill &Mass ratio \dotfill &$0.002425^{+0.000043}_{-0.000041}$\\
~~~~$d/R_*$\dotfill &Separation at mid transit \dotfill &$7.95^{+0.24}_{-0.23}$\\
~~~~$\dot{\gamma}$\dotfill &RV slope$^{1}$ (m/s/day)\dotfill &$-0.00074\pm0.00034$\\
\smallskip\\\multicolumn{2}{l}{Telescope Parameters:}&HIRES\smallskip\\
~~~~$\gamma_{\rm rel}$\dotfill &Relative RV Offset$^{1}$ (m/s)\dotfill &$-11.64\pm0.80$\\
~~~~$\sigma_J$\dotfill &RV Jitter (m/s)\dotfill &$4.26^{+0.45}_{-0.40}$\\
~~~~$\sigma_J^2$\dotfill &RV Jitter Variance \dotfill &$18.1^{+4.1}_{-3.2}$\\
\enddata
\tablenotetext{1}{Reference epoch = 2456802.342883}
\end{deluxetable*}

\begin{figure*}
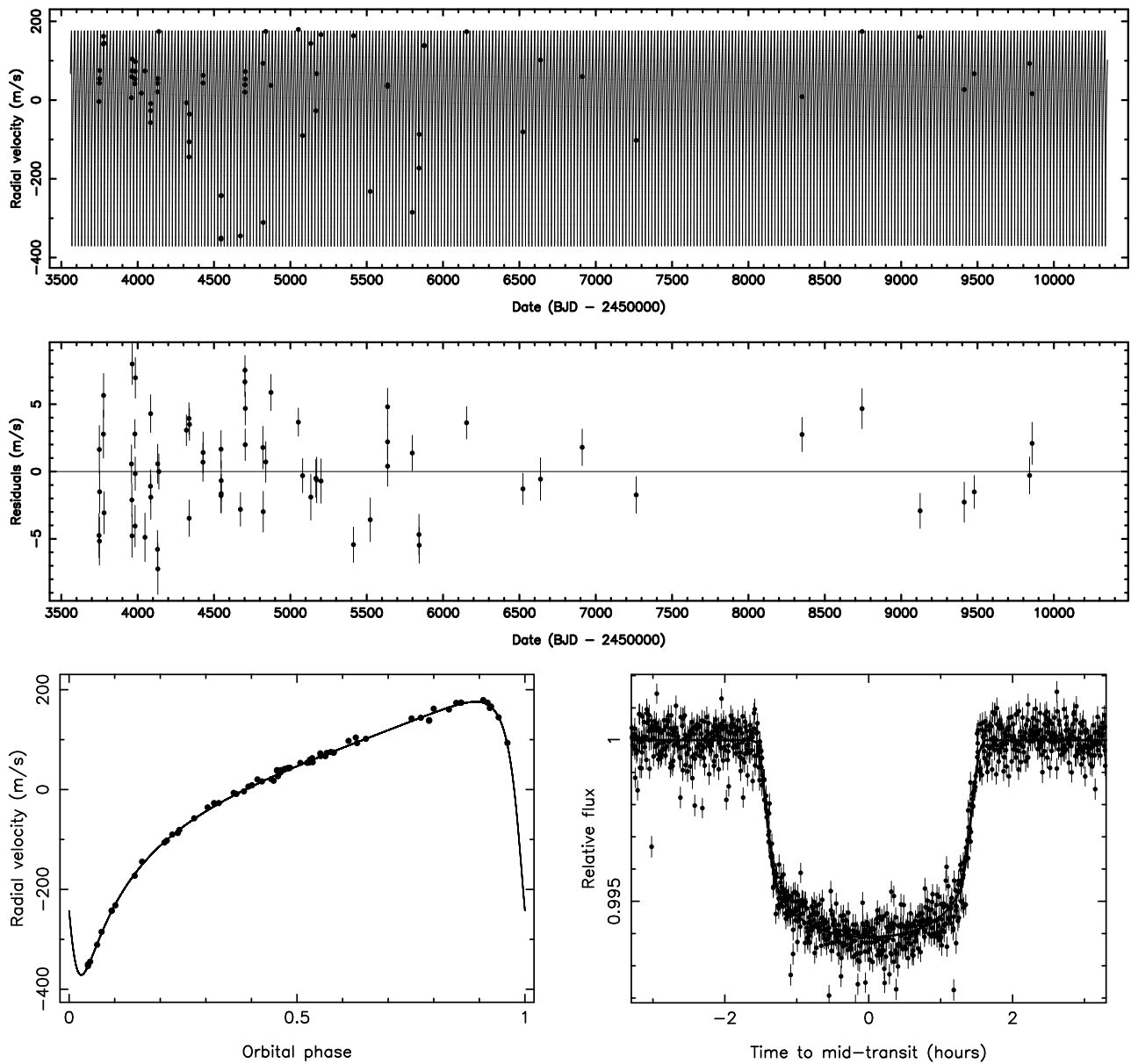

  \begin{center}
    \begin{tabular}{cc}
      \multicolumn{2}{c}{\includegraphics[angle=270,width=17.0cm]{fig_prv.ps}} \\
      \multicolumn{2}{c}{\includegraphics[angle=270,width=17.0cm]{fig_prvres.ps}} \\
      \includegraphics[angle=270,width=8.0cm]{fig_prvfolded.ps} &
      \includegraphics[angle=270,width=8.0cm]{fig_ptran.ps} \\
    \end{tabular}
  \end{center}
  \caption{RV and TESS photometric data for HD~17156. Top panel: All
    Keck/HIRES RV data, spanning a total period of $\sim$17~years,
    along with the best-fit model after applying the {\sc EXOFAST}
    fits described in Section~\ref{system}. Middle panel: Residuals
    from the best-fit model applied to the RV data.  Bottom-left
    panel: RV data folded on the orbital period of the known
    planet. Bottom-right panel: Transit fit from the {\sc EXOFAST}
    analysis to the combined {\it TESS} photometry described in
    Section~\ref{tess}, where all 5 transits have been folded on the
    planetary orbital period.}
  \label{fig:fits}
\end{figure*}

The best-fit RV and transit models are shown in Figure~\ref{fig:fits},
along with their associated data. The top panel shows all of the
Keck/HIRES RV data utilized in this analysis over the full span of
$\sim$17~years. The uncertainties are shown in the plot, but the
median RV uncertainty of 1.415~m/s (see Section~\ref{rv}) is small
compared with the RV semi-amplitude of 274.5~m/s (see
Table~\ref{tab:planet}). The middle panel shows the residuals from the
best-fit model applied to the RV data. The bottom-left panel of
Figure~\ref{fig:fits} shows the RV data folded on the orbital period
provided in Table~\ref{tab:planet}. The bottom-right panel shows the
{\it TESS} photometry from the four sectors described in
Section~\ref{tess} folded on the transit mid-point, along with the
best-fit transit model.


\subsection{Limits on Additional Planets}
\label{limits}

The various data sources described in Section~\ref{obs} provide a
compelling means through which to quantify the presence of other
possible companions within the HD~17156 system. For example, as stated
in Section~\ref{tess}, the RMS scatter of the {\it TESS} photometry is
380~ppm over the sectors for which the target was observed. Adopting
the stellar parameters provided in Table~\ref{tab:star}, this is
equivalent to the transit depth of a 0.3~$R_J$ planet. The transit of
such an additional planet, assuming the orbital inclination is
appropriately aligned, would thus have been detected within the {\it
  TESS} photometry if it occurred during the observed sectors. The
imaging data (Section~\ref{imaging}) demonstrate that there are
unlikely to be stellar companions within the system, and imaging
possible planets rely on a correct assessment of their eccentricity
\citep{kane2013c} and orbital ephemerides \citep{kane2018c,li2021a}.

\begin{figure}
  \includegraphics[width=8.5cm]{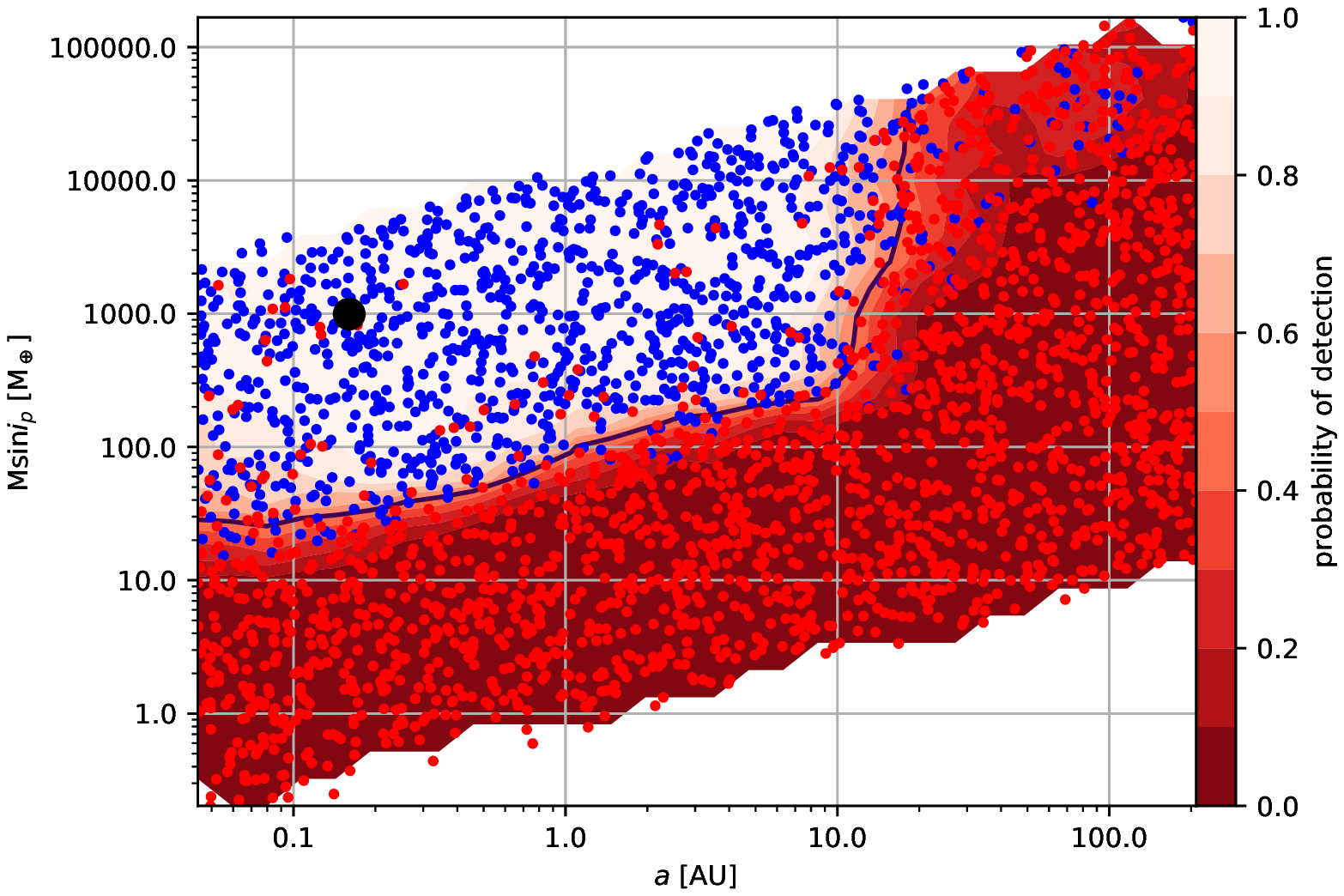}
  \caption{Injection-recovery results that determine the sensitivity
    of the HD~17156 RV data to planetary signatures as a function of
    planetary mass ($M_p \sin i$) and semi-major axis ($a$). The large
    black dot indicates the mass and semi-major axis of the known
    planet. The blue dots represent injected planetary signatures that
    were successfully recovered and the red dots represent those
    planets that were not recovered. The color scale corresponding to
    the probability contours of detecting a planet of a given mass and
    semi-major axis is shown on the right vertical axis.}
  \label{fig:inj}
\end{figure}

The greatest constraints on additional planets arise from the RV data
described in Section~\ref{rv}. Table~\ref{tab:planet} includes the
linear slope of the RV data, incorporated as a free parameter in the
overall fit to the data. The RV data exhibit a negligible slope,
consistent with no further giant planet companions within the
system. To investigate this further, the RV data were used to perform
an injection-recovery test that quantifies the completeness of the
data for the detection of additional planetary signatures, as
described by \citet{howard2016}. The method injects a variety of
planet mass and semi-major axis signatures into the RV data, where the
observation epochs and noise properties of the data are preserved. The
injected signatures assume circular orbits and the fits to the
resulting datasets are performed using the RadVel package
\citep{fulton2018a}. The stellar mass from Table~\ref{tab:star} was
used for translating between the $M_p \sin i$ values and the RV
semi-amplitude.

The injection-recovery results are shown in Figure~\ref{fig:inj} as a
function of planet mass and semi-major axis, where the masses are
provided in Earth masses ($M_\oplus$). The blue dots represent
injected planets that were recovered and the red dots represent those
that were not recovered. The shaded contours provide the probability
of detection for the given planet mass and semi-major axis, indicated
by color scale shown on the right vertical axis. The large black dot
indicates the mass and semi-major axis of HD~17156b, which prominently
lies within the regions of parameter space for which the RV data are
sufficiently sensitive for a successful detection. These results
establish that our RV data are sufficient to rule out additional
planets within the system of Jupiter-mass planets within 10~AU, and of
Saturn-mass planets within 1~AU. Planets below the detection limit may
still be present in the system, including terrestrial planets at a
wide range of separations from the host star, provided that they are
not dynamically excluded by the gravitational influence of the known
eccentric giant planet. Indeed, systems with a single eccentric planet
can serve as excellent RV standards, due to the likely exclusion of
terrestrial planets that would otherwise contribute to the RV error
budget \citep{brewer2020}.


\section{Eccentricity Origin of the Known Planet}
\label{dynamical}

The relatively high eccentricity of HD~17156b poses an interesting
question regarding the dynamical origin of the orbit, particularly as
the calculated tidal circularization timescale of $\sim$20.7~Gyr (see
Table~\ref{tab:planet}) is large compared with the estimated stellar
age of $\sim$3.3~Gyr (see Table~\ref{tab:star}). All of the data
presented in this paper are consistent with a scenario in which there
are no other companions within the HD~17156 system other than the star
and planet. The RV data, described in Section~\ref{rv} and
Section~\ref{limits}, indicate that there are no other stellar or
planetary companions down to the detection limit of the data. There
may be additional terrestrial planets within several AU of the host
star, but such planets are unlikely to have been a major dynamical
contributor to the observed system. The speckle imaging data,
described in Section~\ref{imaging} and Section~\ref{limits}, are also
consistent with a lack of stellar companions within the system,
including wide separation orbits. This is in contrast to several other
known systems that harbor highly eccentric planets, such as
HD~80606/HD~80607 \citep{naef2001b,liu2018b} and HD~20781/HD~20782
\citep{jones2006,mack2014,kane2016b}. Such binary systems present
opportunities for orbital excitation of giant planets via
gravitational interaction with the binary star companion during times
of close approach \citep{malmberg2009,quarles2018a}. In the absence of
stellar companions, as is the case for HD~17156, planet-planet
scattering as the source of high eccentricity is a possible alternate
scenario \citep{chatterjee2008,ford2008c,mustill2017,carrera2019b}.

\begin{figure*}
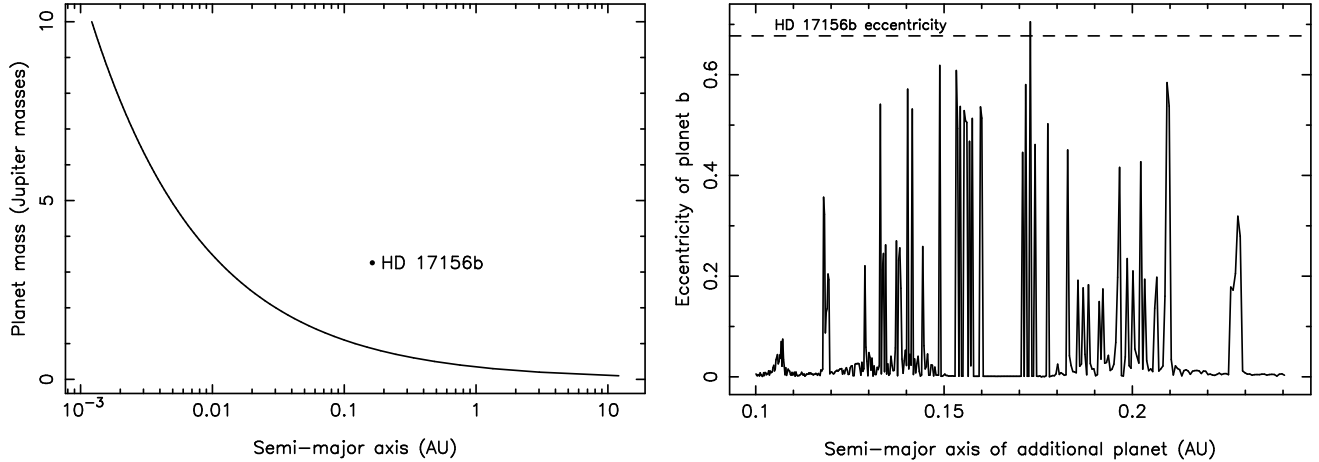

  \centering
  \begin{tabular}{cc}
    \includegraphics[angle=270,width=8.5cm]{fig_amd.ps} &
    \includegraphics[angle=270,width=8.5cm]{fig_ecc.ps}
  \end{tabular}
  \caption{Left: The mass and semi-major axis of an additional planet
    in a circular orbit whose angular momentum equals the angular
    momentum deficit (AMD) for the HD~17156 system. The dot indicated
    the known mass and semi-major axis of HD~17156b. Right:
    Eccentricity of HD~17156b as a function of the semi-major axis for
    the additional planet. The horizontal dashed line indicated the
    current eccentricity of HD~17156b.}
  \label{fig:amd}
\end{figure*}

To investigate planet-planet scattering scenarios for the system, we
calculated the angular momentum deficit (AMD) for the system
\citep{laskar1997}. The AMD describes the difference in total angular
momentum between the eccentric orbits present within a system and
equivalent circular orbits. The AMD thus may indicate lost angular
momentum through planet ejection scenarios, and may also be used as an
indicator of long-term planetary system stability
\citep{laskar2017,he2020c}. The AMD for the HD~17156 system is $3.33
\times 10^{42}$~kg\,m$^2$/s, which is approximately twice the orbital
angular momentum of Uranus. We calculated a range of masses and
semi-major axis values for a planet in a circular orbit that would
have an angular momentum equivalent to the AMD of the HD~17156 system,
the results of which are plotted in the left panel of
Figure~\ref{fig:amd}, along with the location of HD~17156b. These
masses and semi-major axes encompass a broad range of values, and only
a small subset of this full range are expected to result in
significant planet-planet interactions.

We conducted hundreds of dynamical simulations via N-body integrations
using the Mercury Integrator Package \citep{chambers1999}. The
simulations adopted a time resolution of 0.1~days and used a hybrid
symplectic/Bulirsch-Stoer integrator with a Jacobi coordinate system
to provide increased accuracy for close encounters
\citep{wisdom1991,wisdom2006b}. We used the parameters for the known
planet HD~17156b, shown in Table~\ref{tab:planet}, but reduced the
orbital eccentricity to zero. An additional planet was placed in a
circular orbit in the semi-major axis range of 0.1--1.0~AU with a mass
dictated by the AMD calculations shown in the left panel of
Figure~\ref{fig:amd}. These system architectures were used as input
for the dynamical simulations, each of which were executed for a total
duration of $10^6$~years, equivalent to $1.72 \times 10^7$ orbits of
HD~17156b. For each simulation, the final eccentricity of HD~17156b
was recorded.

The results of these simulations are represented in the right panel of
Figure~\ref{fig:amd}, which plots the semi-major axis of the
additional planet and the final eccentricity of planet b. The
horizontal dashed line indicates the current eccentricity of
HD~17156b, $e = 0.6772$, as shown in Table~\ref{tab:planet}. The
majority of simulation cases result either in negligible interactions
between the planets, or the additional planet being lost to the
gravitational well of the host star. The spikes in eccentricity are
the result of planet-planet scattering events in which the additional
planet is ejected from the system, transferring significant angular
momentum to the the remaining planet. In all such cases, the remaining
planet is HD~17156b since it is substantially more massive than the
additional planet, as seen in the left panel of
Figure~\ref{fig:amd}. The outcome of one simulation produced an
eccentricity for HD~17156b of 0.7, slightly higher than its present
value, caused by an additional planet with a mass of 0.84~$M_J$ and
semi-major axis of 0.173~AU. This demonstrates the viability of the
planet-planet scattering scenario as the source of the HD~17156b
eccentricity. Note that this investigation is intended as a
first-order study of possible planet-planet scattering scenarios,
ignoring factors such as interactions with the disk during formation
\citep{clement2021e}, and further planets that may have participated
in the dynamical evolution of the system.


\section{Conclusions}
\label{conclusions}

Planetary system architectures are at the forefront of exoplanetary
science investigations, enabled by the vast amount of statistical data
that are provided by discoveries over recent decades. The origin of
highly eccentric planets, one of the first observed significant
divergences from the solar system architecture, remains an active area
of research. HD~17156b was a key exoplanet detection, since the
discovery of its transit placed it in a separate category from the
population of hot Jupiters that was starting to emerge. The subsequent
data for the system reveal that the star does not appear to have
stellar companions, and the only known planet is alone in the system
down to the detection limit of the data, ruling out additional
Jupiter-mass planets within 10~AU, and of Saturn-mass within 1~AU.

Although there are numerous scenarios that may produce highly
eccentric orbits, such as disk interactions and encounters with a
stellar binary, the evidence in this case points toward a possible
planet-planet scattering event. Although the use of the AMD to
evaluate such planet-planet scattering scenarios is a first-order
investigation tool, it does reveal possible eccentricity
progenitors. There are now many other similar transiting systems with
eccentric orbits that have been detected, thanks largely to the
discoveries of the {\it TESS} mission. As this population continues to
grow, a more exhaustive analysis of the eccentric planet population as
a function of stellar and planetary multiplicity will provide further
insights into the origins of highly eccentric orbits.


\section*{Acknowledgements}

The authors would like to thank Teo Mo\v{c}nik for discussions
regarding HD~17156. G.W.H. acknowledges long-term support from NASA,
NSF, Tennessee State University, and the State of Tennessee through
its Centers of Excellence program. P.D. acknowledges support by a 51
Pegasi b Postdoctoral Fellowship from the Heising-Simons
Foundation. T.F. acknowledges support from the University of
California President's Postdoctoral Fellowship Program. We gratefully
acknowledge the efforts and dedication of the Keck Observatory staff
for support of HIRES and remote observing. We recognize and
acknowledge the cultural role and reverence that the summit of
Maunakea has within the indigenous Hawaiian community. We are deeply
grateful to have the opportunity to conduct observations from this
mountain. Observations in the paper also made use of the
High-Resolution Imaging instrument 'Alopeke. 'Alopeke was funded by
the NASA Exoplanet Exploration Program and built at the NASA Ames
Research Center by Steve B. Howell, Nic Scott, Elliott P. Horch, and
Emmett Quigley. 'Alopeke was mounted on the Gemini North telescope of
the international Gemini Observatory, a program of NSF's NOIRLab,
which is managed by the Association of Universities for Research in
Astronomy (AURA) under a cooperative agreement with the National
Science Foundation on behalf of the Gemini Observatory partnership:
the National Science Foundation (United States), National Research
Council (Canada), Agencia Nacional de Investigaci\'{o}n y Desarrollo
(Chile), Ministerio de Ciencia, Tecnolog\'{i}a e Innovaci\'{o}n
(Argentina), Minist\'{e}rio da Ci\^{e}ncia, Tecnologia,
Inova\c{c}\~{o}es e Comunica\c{c}\~{o}es (Brazil), and Korea Astronomy
and Space Science Institute (Republic of Korea). The results reported
herein benefited from collaborations and/or information exchange
within NASA's Nexus for Exoplanet System Science (NExSS) research
coordination network sponsored by NASA's Science Mission Directorate.


\software{EXOFAST \citep{eastman2013,eastman2020}, Mercury
  \citep{chambers1999}, RadVel \citep{fulton2018a}}




\end{document}